\documentclass[10pt,conference]{IEEEtran}
\usepackage[pass,letterpaper]{geometry}

\usepackage[T1]{fontenc}
\usepackage[utf8]{inputenc}

\usepackage{cite}

\ifCLASSINFOpdf
\else
\fi
\usepackage{graphicx}

\usepackage[cmex10]{amsmath}
\usepackage{array}
\usepackage{booktabs}
\usepackage{amssymb}
\usepackage{url}

\DeclareMathOperator{\rmname}{name}
\DeclareMathOperator{\rmcont}{\pi}
\DeclareMathOperator{\rmsig}{sig}
\DeclareMathOperator{\rmsub}{subtype}
\DeclareMathOperator{\rmsuper}{supertype}
\DeclareMathOperator{\rmsubsuper}{subOrSuper}
\DeclareMathOperator{\rmtype}{type}
\DeclareMathOperator{\rmsim}{sim}
\DeclareMathOperator{\rmsimu}{sim_{p}}

\DeclareMathOperator{\rmcallers}{callers}

\DeclareMathOperator{\tp}{tp}
\DeclareMathOperator{\fp}{fp}
\DeclareMathOperator{\fn}{fn}
\DeclareMathOperator{\precision}{precision}
\DeclareMathOperator{\recall}{recall}

\usepackage[usenames]{color}
\definecolor{darkgray}{RGB}{90,90,90}
\definecolor{lightgray}{RGB}{210,210,210}
\newcommand\xbar[1]{#1 {\color{darkgray} \rule{\dimexpr #1pt * 16}{5.5pt}}{\color{lightgray} \rule{\dimexpr 16pt - (#1pt * 16)}{5.5pt}}}

\usepackage{listings}
\usepackage{lipsum}
\usepackage{courier}
\definecolor{javablue}{rgb}{0.25,0,1} 
\definecolor{javagreen}{rgb}{0.25,0.5,0.35} 
\definecolor{javapurple}{rgb}{0.5,0,0.35} 
\definecolor{javadocblue}{rgb}{0.25,0.35,0.75} 
\begin{document}
%
\title{RefDiff: Detecting Refactorings in Version Histories}


\author{\IEEEauthorblockN{
Danilo Silva\IEEEauthorrefmark{1}, 
Marco Tulio Valente\IEEEauthorrefmark{2}}
\IEEEauthorblockA{Department of Computer Science\\
Universidade Federal de Minas Gerais\\
Belo Horizonte, Brazil\\
Email: \IEEEauthorrefmark{1}danilofs@dcc.ufmg.br, 
\IEEEauthorrefmark{2}mtov@dcc.ufmg.br}
}


%


\maketitle

\begin{abstract}
Refactoring is a well-known technique that is widely adopted by software engineers to improve the design and enable the evolution of a system.
Knowing which refactoring operations were applied in a code change is a valuable information to understand software evolution, adapt software components, merge code changes, and other applications. In this paper, we present RefDiff, an automated approach that identifies refactorings performed between two code revisions in a git repository. RefDiff employs a combination of heuristics based on static analysis and code similarity to detect 13 well-known refactoring types.
In an evaluation using an oracle of 448 known refactoring operations, distributed across seven Java projects, our approach achieved precision of 100\% and recall of 88\%.
Moreover, our evaluation suggests that RefDiff has superior precision and recall than existing state-of-the-art approaches.
\end{abstract}


%
\IEEEpeerreviewmaketitle

\section{Introduction}

Refactoring is a well-known technique to improve the design of a system and enable its evolution~\cite{Fowler:1999}.
In fact, existing studies~\cite{MurphyHill2012, tsantalis_empiricalstudy, Kim:2012:FSE, kim-tse-2014, fse2016-why-we-refactor} present strong evidences that refactoring is frequently applied by development teams, and it is an important aspect of their software maintenance workflow.

Therefore, knowing about the refactoring activity in a code change is a valuable information to help researchers to understand software evolution.
For example, past studies have used such information to shed light on important aspects of refactoring practice, such as: how developers refactor~\cite{MurphyHill2012}, the usage of refactoring tools~\cite{negara2013, MurphyHill2012}, the motivations driving refactoring~\cite{Kim:2012:FSE, kim-tse-2014, fse2016-why-we-refactor}, the risks of refactoring~\cite{Kim:2012:FSE, kim-tse-2014, Kim:2011, weissgerber2006refactorings, bavota2012does}, and the impact of refactoring on code quality metrics~\cite{Kim:2012:FSE, kim-tse-2014}.
Moreover, knowing which refactoring operations were applied in the version history of a system may help in several practical tasks.
For example, in a study by Kim~et~al.~\cite{Kim:2012:FSE}, many developers mentioned the difficulties they face when reviewing or integrating code changes after large refactoring operations, which moves or renames several code elements. Thus, developers feel discouraged to refactor their code. If a tool is able to identify such refactoring operations, it can possibly resolve merge conflicts automatically. 
Moreover, diff visualization tools can also benefit from such information, presenting refactored code elements side-by-side with their corresponding version before the change.
Another application for such information is adapting client code to a refactored version of an API it uses~\cite{henkel2005catchup, Xing:2008:JDevAn}. If we are able to detect the refactorings that were applied to an API, we can replay them on the client code automatically.

Although there are approaches capable of detecting refactorings automatically, there are still some issues that hinder their application. Specifically, the precision and recall of such approaches still need improvements.
In this paper, we try to fill this gap by proposing RefDiff, an automated approach that identifies refactorings performed in the version history of a system.
RefDiff employs a combination of heuristics based on static analysis and code similarity to detect 13 well-known refactoring types.
When compared to existing approaches, RefDiff leverages existing techniques and also introduces some novel ideas, such as the adaptation of the classical TF-IDF similarity measure from information retrieval to compare refactored code elements, and a new strategy to compare the similarity of fields by taking into account the similarity of the statements that reads from or writes to them.

In the paper, we also describe in details a study to evaluate the precision and recall of RefDiff and three existing refactoring detection approaches: Refactoring Miner~\cite{fse2016-why-we-refactor}, Refactoring Crawler~\cite{dig2006automated}, and Ref-Finder~\cite{prete2010template,Kim:2010:RefFinder}. In our study, RefDiff achieved precision of 100\% and recall of 88\%, which were the best results among the evaluated approaches.

In summary, the contributions we deliver in this work are:
\begin{itemize}
\item RefDiff, which is a new approach to detect refactoring in version histories.
We provide a publicly available\footnote{RefDiff and all evaluation data are public available in GitHub:\\
\url{https://github.com/aserg-ufmg/RefDiff}} implementation of our approach that is capable of finding refactorings in Java code within git repositories in a fully automated way;
\item a publicly available oracle of 448 known refactoring operations, applied to seven Java systems, that serves as an evaluation benchmark for refactoring detection approaches; and
\item an evaluation of the precision and recall of RefDiff, comparing it with three state-of-the-art approaches.
\end{itemize}

The remainder of this paper is structured as follows. Section~\ref{SecBackground} describes related work, focusing on the three approaches we compare with RefDiff. Section~\ref{SecApproach} presents the proposed approach in details.
Section~\ref{SecEval} describes how we evaluated RefDiff and discusses the achieved results.
Section~\ref{SecThreats} discusses threats to validity and we conclude the paper in Section~\ref{SecConclusion}.

\section{Related Work}
\label{SecBackground}

Empirical studies on refactoring rely on means to identify refactoring activity. Thus, many different techniques have been proposed and employed for this task.
For example, Murphy-Hill~et~al.~\cite{MurphyHill2012} collected refactoring usage data using a framework that monitors user actions in the Eclipse IDE, including calls to refactoring commands.
Negara~et~al.~\cite{negara2013} also used the strategy of instrumenting the IDE to infer refactorings from fine-grained code edits.
Other studies use metadata from version control systems to identify refactoring changes. For example, Ratzinger~et~al.~\cite{ratzinger2008relation} search for a predefined set of terms in commit messages to classify them as refactoring changes. In specific scenarios, a branch may be created exclusively to refactor the code, as reported by Kim et al.~\cite{kim-tse-2014}.
Another strategy is employed by Soares et al.~\cite{soares2010making}. They propose an approach that identify behavior-preserving changes by automatically generating and running test-cases. While their approach is intended to guarantee the correct behavior of a system after refactoring, it may also be employed to classify commits as behavior-preserving.
Moreover, many existing approaches are based on static analysis.
This is the case of the approach proposed by Demeyer et al.~\cite{demeyer2000finding}, which finds refactored elements by observing changes in code metrics.

Static analysis is also frequently used to find differences in the source code~\cite{dig2006automated, weissgerber2006identifying, tsantalis_empiricalstudy,prete2010template,Kim:2010:RefFinder}.
Approaches based on comparing source code differences have the advantage of beeing able to identify each refactoring operation performed. As RefDiff is one of these approaches, it can be directly compared with others within this category. In the next sections, we will describe three of such approaches.

\subsection{Refactoring Miner}

Refactoring Miner is an approach introduced by Tsantalis~et~al.~\cite{tsantalis_empiricalstudy}, that was later extend by Silva~et~al.~\cite{fse2016-why-we-refactor} to mine refactorings in large scale in git repositories. This tool is capable of identifying 14 high-level refactoring types: \emph{Rename Package/Class/Method}, \emph{Move Class/Method/Field}, \emph{Pull Up Method/Field}, \emph{Push Down Method/Field}, \emph{Extract Method}, \emph{Inline Method}, 
and \emph{Extract Superclass/Interface}.

Refactoring Miner runs a lightweight algorithm, similar to the UMLDiff proposed by Xing and Stroulia~\cite{Xing:2005}, for differencing object-oriented models, inferring the set of classes, methods, and fields added, deleted or moved between two code revisions. 
First, the algorithm matches code entities in a top-down order (starting from the classes and going to the methods and fields) looking for exact matches on their names and signatures (in the case of methods).
Next, the removed/added elements between the two models are matched based only on the equality of their names in order to find changes in the signatures of fields and methods.
Third, the removed/added classes are matched based on the similarity of their members at signature level.
Finally, a set of rules enforcing structural constraints is applied to identify specific types of refactorings.

In a first study, using the version histories of JUnit, HTTPCore, and HTTPClient, Tsantalis~et~al.~\cite{tsantalis_empiricalstudy} found 8 false positives for the \emph{Extract Method} refactoring (96.4\% precision) and 4 false positives for the \emph{Rename Class} refactoring (97.6\% precision). No false positives were found for the remaining refactorings.
In a second study that mined refactorings in 285 GitHub hosted Java repositories, Silva~et~al.~\cite{fse2016-why-we-refactor} found 1,030 false positives out of 2,441 refactorings (63\% precision). However, the authors also evaluated Refactoring Miner using as a benchmark the dataset reported by Chaparro~et~al.~\cite{Chaparro:2014}, in which it achieved 93\% precision and 98\% recall.

\subsection{Refactoring Crawler}

Refactoring Crawler, proposed by Dig~et~al.~\cite{dig2006automated}, is an approach capable of finding seven high-level refactoring types: \emph{Rename Package/Class/Method}, \emph{Pull Up Method}, \emph{Push Down Method}, \emph{Move Method}, and \emph{Change Method Signature}.
It uses a combination of a syntactic analysis to detect refactoring candidates and a more expensive reference graph analysis to refine the results.

First, Refactoring Crawler analyzes the abstract syntax tree of a program and produces a tree, in which each node represents a source code entity (package, class, method, or field).
Then, it employs a technique known as \emph{shingles encoding} to find 
similar pairs of entities, which are candidates for refactorings.
Shingles are representations for strings with the following property: if a string changes slightly, then its shingles also change slightly.
In a second phase, Refactoring Crawler applies specific strategies for detecting each refactoring type, and computes a more costly metric that determines the similarity of references among code entities in the two versions of the system. For example, two methods are similar if the sets of methods that call them are similar, and the sets of methods they call are also similar.
The strategies to detect refactorings are repeated in a loop until no new refactorings are found. Therefore, the detection of a refactoring, such as a rename, may change the reference graph of code elements and enable the detection of new refactorings.

The authors evaluated Refactoring Crawler comparing pairs of releases of three open source software components: Eclipse UI, Struts, and JHotDraw. Such components were chosen because they provided detailed release notes describing API changes. The authors relied on such information and on manual inspection to build an oracle of known refactorings in those releases, containing 131 refactorings in total.
The reported results are: Eclipse UI (90\% precision and 86\% recall), Struts (100\% precision and 86\% recall), and JHotDraw (100\% precision and 100\% recall).

\subsection{Ref-Finder}

Ref-Finder, proposed by Prete~et~al.~\cite{prete2010template,Kim:2010:RefFinder}, is an approach based on logic programming capable of identifying 63 refactoring types from the Fowler's catalog\cite{Fowler:1999}.
The authors express each refactoring type by defining structural constraints, before and after applying a refactoring to a program, in terms of template logic rules.

First, Ref-Finder traverses the abstract syntax tree of a program and extracts facts about code elements, structural dependencies, and the content of code elements, to represent the program in terms of a database of logic facts. Then, it uses a logic programming engine to infer concrete refactoring instances, by creating a logic query based on the constraints defined for each refactoring type.
The definition of refactoring types also consider ordering dependencies among them. This way, lower-level refactorings may be queried to identify higher-level, composite refactorings.
The detection of some types of refactoring requires a special logic predicate that indicates that the similarity between two methods is above a threshold. For this purpose, the authors implemented a block-level clone detection technique, which removes any beginning and trailing parenthesis, escape characters, white spaces and return keywords and computes word-level similarity between the two texts using the longest common sub-sequence algorithm.

The authors evaluated Ref-Finder in two case studies.
In the first one, they used code examples from the Fowler's catalog to create instances of the 63 refactoring types. The authors reported 93.7\% recall and 97.0\% precision for this first study.
In the second study, the authors used three open-source projects: Carol, jEdit, and Columba. In this case, Ref-Finder was executed in randomly selected pairs of versions. From the 774 refactoring instances found, the authors manually inspected a sample of 344 instances and found that 254 were correct (73.8\% precision).
However, in a study by Soares~et~al.~\cite{Soares:2013} using a set of randomly select versions of JHotDraw and Apache Common Collections containing 81 refactoring instances in total, Ref-Finder achieved only 35\% precision and 24\% recall.

\section{Proposed Refactoring Detection Algorithm}
\label{SecApproach}

RefDiff employs a combination of heuristics based on static analysis and code similarity to detect refactorings between two revisions of a system.
Thus, RefDiff takes as input two versions of a system, and outputs a list of refactorings found.

The detection algorithm is divided in two main phases: Source Code Analysis and Relationship Analysis.
In the first phase, the source code of the system is parsed and analyzed to build a model that represents each high level source code entity, such as types, methods, and fields.
Two models are built to represent the system before ($E_b$) and after the changes ($E_a$).
For efficiency, only code entities that belong to modified source files (added, removed or edited) are analyzed.
Each of these two models is a set of types, method, and fields contained in the source code. Specifically, $E_b = (T_b \cup M_b \cup F_b)$, such that $T_b$, $M_b$, and $F_b$ are the sets of types, methods, and fields in the source code before the changes, and $E_a = (T_a \cup M_a \cup F_a)$, such that $T_a$, $M_a$, and $F_a$ are the sets of types, methods, and fields after the changes.

The second phase of the algorithm, Relationship Analysis, consists in finding relationships between source code entities before and after the code changes.
Specifically, the algorithm builds a bipartite graph with two sets of vertices: code entities before ($E_b$) and code entities after ($E_a$).
The edges of this graph are represented by the set of relationships $R$ between code entities.
For example, a certain method $m_1 \in M_b$ may correspond to a method $m_2 \in M_a$ that was renamed by a developer. This would correspond to a \emph{Rename Method} relationship between $m_1$ and $m_2$ and, consequently, to a \emph{Rename Method} refactoring.

Table~\ref{TabRelationships} presents all relationships that RefDiff can identify between types, methods, or fields. We search for relationships between source code entities considering each relationship type in the order they are presented in the table. The following sections detail how such relationships are identified.

\begin{table*}[htb]
\renewcommand{\arraystretch}{1.2}
\setlength{\tabcolsep}{4pt}
\caption{Relationship types}
\label{TabRelationships}
\centering
\footnotesize
\begin{tabular}{@{}lll@{}}
\toprule
Relationship & \multicolumn{2}{@{}l}{Condition} \\
\midrule
& \multicolumn{2}{@{}l}{$(t_b, t_a) \in T_b \times T_a$, such that:}\\
Same Type & & $\rmname(t_b) = \rmname(t_a) \land \rmcont(t_b) \sim \rmcont(t_a)$\\
Rename Type & & $\rmname(t_b) \neq \rmname(t_a) \land \rmcont(t_b) \sim \rmcont(t_a) \land \rmsim(t_b, t_a) > \tau$\\
Move Type   & & $\rmname(t_b) = \rmname(t_a) \land \rmcont(t_b) \nsim \rmcont(t_a) \land \rmsim(t_b, t_a) > \tau$\\
Move and Rename Type & & $\rmname(t_b) \neq \rmname(t_a) \land \rmcont(t_b) \nsim \rmcont(t_a) \land \rmsim(t_b, t_a) > \tau$\\
Extract Supertype & & $(\nexists x \in T_b\,|\,x \sim t_a) \land  (\exists y \in T_a\,|\,t_b \sim y \land \rmsub(y, t_a)) \land \rmsimu(t_a, t_b) > \tau$\\
\addlinespace
& \multicolumn{2}{@{}l}{$(m_b, m_a) \in M_b \times M_a$, such that:}\\
Same Method & & $\rmsig(m_b) = \rmsig(m_a) \land \rmcont(m_b) \sim \rmcont(m_a)$\\
Rename Method & & $\rmname(m_b) \neq \rmname(m_a) \land \rmcont(m_b) \sim \rmcont(m_a) \land \rmsim(m_b, m_a) > \tau$\\
Change Method Signature & & $\rmname(m_b) = \rmname(m_a) \land \rmsig(m_b) \neq \rmsig(m_a) \land \rmcont(m_b) \sim \rmcont(m_a) \land \rmsim(m_b, m_a) > \tau$\\
Pull Up Method & & $\rmsig(m_b) = \rmsig(m_a) \land \rmsub(\rmcont(m_b)^{\sim}, \rmcont(m_a)) \land \rmsim(m_b, m_a) > \tau$\\
Push Down Method & & $\rmsig(m_b) = \rmsig(m_a) \land \rmsuper(\rmcont(m_b)^{\sim}, \rmcont(m_a)) \land \rmsim(m_b, m_a) > \tau$\\
Move Method & & $\rmname(m_b) = \rmname(m_a) \land \rmcont(m_b) \nsim \rmcont(m_a) \land \neg \rmsubsuper(\rmcont(m_b)^{\sim}, \rmcont(m_a)) \land \rmsim(m_b, m_a) > \tau$\\
Extract Method & & $(\nexists x \in M_b\,|\,x \sim m_a) \land (\exists y \in M_a\,|\,m_b \sim y \land y \in \rmcallers(m_a)) \land \rmsimu(m_a, m_b) > \tau$\\
Inline Method & & $(\nexists x \in M_a\,|\,m_b \sim x) \land (\exists y \in M_b\,|\,y \sim m_a \land y \in \rmcallers(m_b)) \land \rmsimu(m_b, m_a) > \tau$\\
\addlinespace
& \multicolumn{2}{@{}l}{$(f_b, f_a) \in F_b \times F_a$, such that:}\\
Same Field & & $\rmname(f_b) = \rmname(f_a) \land \rmtype(f_b) = \rmtype(f_a) \land \rmcont(f_b) \sim \rmcont(f_a)$\\
Pull Up Field & & $\rmname(f_b) = \rmname(f_a) \land \rmtype(f_b) = \rmtype(f_a) \land \rmsub(\rmcont(f_b)^{\sim}, \rmcont(f_a)) \land \rmsim(f_b, f_a) > \tau$\\
Push Down Field & & $\rmname(f_b) = \rmname(f_a) \land \rmtype(f_b) = \rmtype(f_a) \land \rmsuper(\rmcont(f_b)^{\sim}, \rmcont(f_a)) \land \rmsim(f_b, f_a) > \tau$\\
Move Field & & $\rmname(f_b) = \rmname(f_a) \land \rmtype(f_b) = \rmtype(f_a) \land \rmcont(f_b) \nsim \rmcont(f_a) \land \neg \rmsubsuper(\rmcont(f_b)^{\sim}, \rmcont(f_a)) \land \rmsim(f_b, f_a) > \tau$\\
\bottomrule
\end{tabular}

\vspace{1em}
\begin{tabular}{@{}lll@{}}
\midrule
\begin{tabular}{@{}ll@{}}
$\rmname(e)$ & simple name of a code entity $e$\\
$\rmsig(m)$ & signature of a method $m$\\
$\rmtype(f)$ & type of a field $f$\\
$\rmsub(t_1, t_2)$ & $t_1$ is subtype of $t_2$\\
$\rmsuper(t_1, t_2)$ & $t_1$ is supertype of $t_2$\\
$\rmsubsuper(t_1, t_2)$ & $\rmsub(t_1, t_2) \lor \rmsuper(t_1, t_2)$\\
\end{tabular}
& &
\begin{tabular}{@{}ll@{}}
$\rmcont(e)$ & container entity of a code entity $e$ (it may be a type or a package)\\
$e_1 \sim e_2$ & exists a matching relationship between $e_1$ and $e_2$\\
$e_1 \nsim e_2$ & does not exists a matching relationship between $e_1$ and $e_2$\\
$e^{\sim}$ & the code entity that matches with $e$ after the change\\
$\rmcallers(m_a)$ & the set of methods that call $m_a$\\
$\rmsim(e_1, e_2)$ & similarity index between $e_1$ and $e_2$\\
$\rmsimu(e_1, e_2)$ & similarity index between $e_1$ and $e_2$ for non-matching relationships\\
\end{tabular}
\\
\midrule
\end{tabular}
\end{table*}

\subsection{Matching Relationships}
\label{SecMatchingRelationships}

Some kinds of relationships map code entities before the change to code entities after the change. For example, let $t_1 \in T_b$ be a type in the version before the change. If our algorithm finds another type $t_2 \in T_a$ with the same qualified name, it adds a relationship \emph{Same Type} between $t_1$ and $t_2$ in $R$. This is a matching relationship, because $t_1$ corresponds to $t_2$ after the change. Other examples of matching relationship are \emph{Move Type}, \emph{Rename Type}, and \emph{Pull Up Method}.
In contrast, suppose that our algorithm finds that $m_2$ is a method that was extracted from another method $m_1$. In this case, there is an \emph{Extract Method} relationship between $m_1$ and $m_2$, but this is not a matching relationship, because $m_1$ does not correspond to $m_2$ after the change. From this point on, we use the notation ${e_1 \sim e_2}$ to represent a matching relationship between $e_1$~and~$e_2$.

We discriminate matching relationships from non-matching relationships because their detection algorithm is similar.
For each matching relationship type, we find all pairs of entities $(e_b, e_a) \in E_b \times E_a$ that fall under the conditions specified in Table~\ref{TabRelationships}. Each relationship type has its specific conditions. For example, as presented in Table~\ref{TabRelationships}, the conditions for identifying a \emph{Rename Method} between $m_1 \in M_b$ and $m_2 \in M_a$ are:
\begin{itemize}
\item the names of $m_1$ and $m_2$ should be different;
\item there should exist a matching relationship between the container classes of $m_1$ and $m_2$; and
\item the similarity index between $m_1$ and $m_2$, denoted by $\rmsim(m_1, m_2)$, should be greater than a threshold $\tau$.
\end{itemize}
Whenever these conditions hold, we add the triple $(e_b, e_a, \rmsim(e_b, e_a))$ in a list of potential \emph{Rename Method} relationships.

The last step to find the actual relationships consists in selecting non-conflicting relationships from the list of potential relationships and add them to the graph. For example, there may be in the list two potential \emph{Rename Method} relationships: $(e_1, e_2, 0.5)$ and $(e_1, e_3, 0.8)$. However, a code entity can not be involved in more than one matching relationship. Thus, only one of them must be chosen, because $e_1$ could not be renamed to $e_2$ and to $e_3$. The criterion we use is to choose the triple with the higher similarity index. This means that, in the aforementioned example, we would choose the triple $(e_1, e_3, 0.8)$ and discard $(e_1, e_2, 0.5)$. In Section~\ref{SecSimilarity} we describe in details how the similarity index is computed.

\subsection{Non-matching Relationships}
\label{SecExtractInlineRelationships}

In the previous section, we discussed that an entity could not be involved in multiple matching relationships, but this property does not hold for non-matching relationships. For example, suppose that a developer extracted some code from a method $m_1$ into a new method $m_2$, i.e., an \emph{Extract Method} refactoring was applied. It is also possible that the developer extracted another part of $m_1$ into a new method $m_3$.

Given that non-matching relationships do not conflict with each other, the algorithm to identify them is simpler. We just need to find all pairs of entities $(e_b, e_a) \in E_b \times E_a$ that fall under the conditions specified in Table~\ref{TabRelationships}. For example, the conditions for identifying an \emph{Extract Method} relationship between $m_1 \in M_b$ and $m_2 \in M_a$ are:
\begin{itemize}
\item there should not exist a method $x \in M_b$ such that $x \sim m_2$ (i.e., $m_2$ was added);
\item there should exist a method $y \in M_a$ such that $m_1 \sim y$ (i.e., $m_1$ was not removed);
\item $y$ should call $m_2$; and
\item the similarity index between $m_2$ and $m_1$, denoted by $\rmsimu(m_2, m_1)$, should be greater than a threshold $\tau$.
\end{itemize}

Besides \emph{Extract Method}, our approach supports the detection of \emph{Inline Method} and \emph{Extract Supertype} relationships.

\subsection{Computing Similarity}
\label{SecSimilarity}

\begin{figure*}[htb]
\renewcommand{\arraystretch}{1.3}
\centering
\footnotesize
\begin{tabular}{@{}llll@{}}
\begin{tabular}{p{6.5cm}}
\multicolumn{1}{c}{\textbf{Source code of a class}} \\
\begin{lstlisting}
public class Calculator {

  public int sum(int x, int y) {
    return x + y;
  }

  public int min(int x, int y) {
    if (x < y) return x;
    else return y;
  }

  public double power(int b, int e) {
    return Math.pow(b, e);
  }
}
\end{lstlisting}\\

\end{tabular}
& {\Large $\Rightarrow$} &
\begin{tabular}{|l|r|r|r|}
\multicolumn{4}{c}{\textbf{Multiset of tokens for each method}} \\
\hline
Token $t$ & $m_{\mathtt{sum}}(t)$ & $m_{\mathtt{min}}(t)$ & $m_{\mathtt{power}}(t)$\\
\hline
$\mathtt{return}$ & 1 & 2 & 1 \\
$\mathtt{x}$      & 1 & 2 & 0 \\
$\mathtt{+}$      & 1 & 0 & 0 \\
$\mathtt{y}$      & 1 & 2 & 0 \\
$\mathtt{;}$      & 1 & 2 & 1 \\
$\mathtt{if}$     & 0 & 1 & 0 \\
$\mathtt{(}$      & 0 & 1 & 1 \\
$\mathtt{<}$      & 0 & 1 & 0 \\
$\mathtt{)}$      & 0 & 1 & 1 \\
$\mathtt{else}$   & 0 & 1 & 0 \\
$\mathtt{Math}$   & 0 & 0 & 1 \\
$\mathtt{.}$      & 0 & 0 & 1 \\
$\mathtt{pow}$    & 0 & 0 & 1 \\
$\mathtt{b}$      & 0 & 0 & 1 \\
$\mathtt{,}$      & 0 & 0 & 1 \\
$\mathtt{e}$      & 0 & 0 & 1 \\
\hline
\end{tabular} &
\begin{tabular}{|r|}
\multicolumn{1}{c}{} \\
\hline
$n_t$\\
\hline
3 \\
2 \\
1 \\
2 \\
3 \\
1 \\
2 \\
1 \\
2 \\
1 \\
1 \\
1 \\
1 \\
1 \\
1 \\
1 \\
\hline
\end{tabular}
\end{tabular}
\caption{Transformation of the body of each method into a multiset of tokens}
\label{fig:FigSourceCodeTransformation}
\end{figure*}

A key element of our algorithm to find relationships, as mentioned previously, is computing the similarity between entities.
The first step to compute similarity of code entities is to represent their source code as a multiset (or bag) of tokens.
A multiset is a generalization of the concept of a set, but it allows multiple instances of the same element.
The multiplicity of an element is the number of occurrences of that element within the multiset. Formally, a multiset can be defined in terms of a multiplicity function $m: U \to \mathbb{N}$, where $U$ is the set of all possible elements. In other words, $m(t)$ is the multiplicity of the element $t$ in the multiset. Note that the multiplicity of an element that is not in the multiset is zero.

For example, Figure~\ref{fig:FigSourceCodeTransformation} depicts the transformation of the source code of three methods ($\mathtt{sum}$, $\mathtt{min}$, and $\mathtt{power}$), of the class $\mathtt{Calculator}$, into multisets of tokens. In the Figure, the multiplicity function $m$ for each method is represented in a tabular form. For example, the multiplicity of the token $\mathtt{y}$ in method $\mathtt{min}$ is two (i.e., $m_{\mathtt{min}}(\mathtt{y}) = 2$), whilst the multiplicity of the token $\mathtt{if}$ in method $\mathtt{power}$ is zero (i.e., $m_{\mathtt{power}}(\mathtt{if}) = 0$).

Later, to compute the similarity between two source code entities $e_1$ and $e_2$, we use a generalization of the Jaccard coefficient, known as weighted Jaccard coefficient~\cite{chierichetti2010finding}.
Let $U$ be the set of all possible tokens and $w(e, t)$ be a weight function of a token $t$ for the entity $e$.
We define the similarity between $e_1$ and $e_2$ by the following formula:

\begin{align}
\rmsim(e_1, e_2) = \frac{\sum_{t \in U} \min(w(e_1, t), w(e_2, t))}
                        {\sum_{t \in U} \max(w(e_1, t), w(e_2, t))}
\end{align}

\subsubsection{Weight of a token for a code entity}

Our similarity function is based on a weighting function $w(e, t)$ that expresses the importance a token $t$ for a code entity $e$.
In fact, some tokens are more important than others to discriminate a code element.
For example, in Figure~\ref{fig:FigSourceCodeTransformation}, all three methods contain the token $\mathtt{return}$. In contrast, only one method ($\mathtt{power}$) contains the token $\mathtt{Math}$. Therefore, the later is a better indicator of similarity between methods than the former.

In order to take this into account, we employ a variation of the TF-IDF weighting scheme~\cite{salton1986introduction}, which is a well-known technique from information retrieval.
TF-IDF, which is the short form of \emph{Term Frequency–Inverse Document Frequency}, reflects how important a term is to a document within a collection of documents.
In the context of code entities, we consider a token as a term, and the body of a method (or class) as a document.

Let $E$ be the set of all code entities and $n_t$ be the number of entities in $E$ that contains the token $t$,
we define the weight of $t$ for a code entity $e$ as the function $w(e, t)$, which is defined by the following formula:
\begin{align}
w(e, t) = m_e(t) \times \mathit{idf}(t)
\end{align}

\noindent where $m_e(t)$ is the multiplicity of $t$ in $e$, and $\mathit{idf}(t)$ is the Inverse Document Frequency, which is defined as:
\begin{align}
\mathit{idf}(t) = \log (1 + \frac{|E|}{n_t})
\end{align}

Note that the value of $\mathit{idf}(t)$ decreases as $n_t$ increases, because the more frequent a token is among the collection of code entities, the less important it is to distinguish code elements.
For example, in Figure~\ref{fig:FigSourceCodeTransformation}, the token $\mathtt{y}$ occurs in two methods ($\mathtt{sum}$ and $\mathtt{min}$). Thus, its $\mathit{idf}$ is:

\[
\mathit{idf}(\mathtt{y}) = 
\log (1 + \frac{|E|}{n_t}) = 
\log (1 + \frac{3}{2}) = 0.398
\]

On the other hand, the token $\mathtt{else}$ occurs in one method ($\mathtt{min}$), and its $\mathit{idf}$ is:

\[
\mathit{idf}(\mathtt{else}) = 
\log (1 + \frac{|E|}{n_t}) = 
\log (1 + \frac{3}{1}) = 0.602
\]


\subsubsection{Similarity of fields}

The similarity of types and methods can be directly computed by the aforementioned similarity function by scanning the source code within their bodies and building the multiset of tokens.
However, such strategy is not suitable to compute the similarity of fields, because they do not have a body. 
To address this limitation, we defined the concept of a virtual body of a field, which is composed of all statements that access the field (read or write) found in the source code of the system. Thus, we are able to compute the multiset of tokens for a field $f_1$ by extracting the tokens of all statements that access $f_1$.
The rationale behind such strategy is that if a field $f_1$ corresponds to a field $f_2$ after a change, the statements that directly used $f_1$ should use $f_2$ after the change, thus, they would likely be similar.

\subsubsection{Similarity for non-matching relationships}

While the similarity function presented previously is suitable to compute whether the source code of two entities are similar, in some situations we need to assess whether the source code of an entity is contained within another one. This is the case of \emph{Extract Supertype}, \emph{Extract Method}, and \emph{Inline Method} relationships. For example, if a method $m_2$ is extracted from $m_1$, the source code of $m_2$ should be contained within $m_1$ prior to the refactoring, although $m_1$ and $m_2$ may be significantly different from each other. Analogously, if a method $m_1$ is inlined into $m_2$, the source code of $m_1$ should be contained within $m_2$.
Thus, we defined a specialized version of the similarity function $\rmsimu$ defined by the following formula:
\begin{align}
\rmsimu(e_1, e_2) = \frac{\sum_{t \in U} \min(w(e_1, t), w(e_2, t))}
                        {\sum_{t \in U} w(e_1, t)}
\end{align}

The rationale behind this formula is that the similarity is at maximum when the multiset of tokens of $e_1$ is a subset of the multiset of tokens of $e_2$, i.e., all tokens from $e_1$ can be found in $e_2$. Note that, given this definition, $\rmsimu(e_1, e_2) \neq \rmsimu(e_2, e_1)$.

\subsection{Calibration of similarity thresholds}
\label{SecCalibration}

Our algorithm relies on thresholds to find relationships between entities, as discussed in Section~\ref{SecMatchingRelationships} and Section~\ref{SecExtractInlineRelationships}. Specifically, for each relationship type, we define a threshold $\tau$ for the minimum similarity that the involved entities should have so that we consider them as a potential relationship. Therefore, the thresholds we choose may affect the precision an recall of the algorithm.
For this reason, we selected such thresholds by applying a well-defined calibration process.

First, we randomly selected a set of ten commits that contain refactorings, in ten different projects (see Table~\ref{TabProjectsCalibration}), drawn from a public dataset used to investigate the reasons for refactoring operations~\cite{fse2016-why-we-refactor}.
We ensured that every refactoring type is covered by at least one commit. All refactorings reported in those commits were initially added to an oracle of known refactorings. Later, for each refactoring type, we run our algorithm using different thresholds values, ranging from $0.1$ to $0.9$ by $0.1$ increments. The output of our algorithm (i.e., the refactorings found) were then compared to the known refactorings from the oracle. Refactorings contained in our oracle were initially classified as true positives, whilst refactorings not contained in our oracle were classified as false positives. Moreover, refactorings in our oracle that were not found were classified as false negatives.
In a second pass, the false positive refactorings were manually inspected to find potential true positives incorrectly classified. This step was necessary because the oracle obtained in the first step may not contain all refactorings in a commit.

\begin{table}[t]
\renewcommand{\arraystretch}{1.2}
\setlength{\tabcolsep}{4pt}
\caption{Projects/commits used int the calibration}
\label{TabProjectsCalibration}
\centering
\begin{tabular}{@{}ll@{}}
\toprule
Repository URL & Commit \\
\midrule
\url{github.com/linkedin/rest.li}   & 54fa890\\
\url{github.com/droolsjbpm/jbpm}    & 3815f29\\
\url{github.com/gradle/gradle}  & 44aab62\\
\url{github.com/jenkinsci/workflow-plugin}  & d0e374c\\
\url{github.com/spring-projects/spring-roo} & 0bb4cca\\
\url{github.com/BuildCraft/BuildCraft}  & a5cdd8c\\
\url{github.com/droolsjbpm/drools}  & 1bf2875\\
\url{github.com/jersey/jersey}  & d94ca2b\\
\url{github.com/undertow-io/undertow}   & d5b2bb8\\
\url{github.com/kuujo/copycat}  & 19a49f8\\
\bottomrule
\end{tabular}

\end{table}

Last, we selected the threshold value by choosing the one that yields the best compromise between precision and recall. Specifically, we choose the value that optimize the $F_1$ score, which is defined as:

\begin{align}
F_1 = 2 \times \frac{\precision \times \recall}{\precision + \recall}
\end{align}

\noindent where $\precision$ and $\recall$ are respectively:

\begin{align}
\precision = \frac{\tp}{\tp + \fp} && \recall = \frac{\tp}{\tp + \fn}
\end{align}

The set of threshold values chosen are presented in the third column of Table~\ref{TabCalibration} ($\tau$).
It is worth noting that the threshold calibration for each relationship type was performed in the order presented in Table~\ref{TabCalibration}, to minimize the effect of dependencies between different relationship types.
For example, suppose that an existing \emph{Move Class} refactoring is not identified. It is likely that false positive \emph{Move Method} instances will be reported, because there should be some similar (or identical) methods when comparing the moved class with itself after the move operation.
Therefore, it is important to calibrate the \emph{Move Type} threshold before  the \emph{Move Method} threshold.

\begin{table}[htbp]
\renewcommand{\arraystretch}{1.2}
\setlength{\tabcolsep}{4pt}
\caption{Thresholds calibration results}
\label{TabCalibration}
\centering
\begin{tabular}{@{}lrrrrrcc@{}}
\toprule
Ref. Type & \# & $\tau$ & TP & FP & FN & Precision & Recall\\
\midrule
Rename Type & 2 & 0.4 & 2 & 0 & 0 & \xbar{1.000} & \xbar{1.000}\\
Move Type & 2 & 0.9 & 2 & 0 & 0 & \xbar{1.000} & \xbar{1.000}\\
Extract Superclass & 2 & 0.8 & 2 & 0 & 0 & \xbar{1.000} & \xbar{1.000}\\
Rename Method & 24 & 0.3 & 22 & 3 & 2 & \xbar{0.880} & \xbar{0.917}\\
Pull Up Method & 7 & 0.4 & 7 & 0 & 0 & \xbar{1.000} & \xbar{1.000}\\
Push Down Method & 2 & 0.6 & 2 & 0 & 0 & \xbar{1.000} & \xbar{1.000}\\
Move Method & 24 & 0.4 & 21 & 1 & 3 & \xbar{0.955} & \xbar{0.875}\\
Extract Method & 25 & 0.1 & 25 & 9 & 0 & \xbar{0.735} & \xbar{1.000}\\
Inline Method & 6 & 0.3 & 5 & 2 & 1 & \xbar{0.714} & \xbar{0.833}\\
Pull Up Field & 2 & 0.5 & 2 & 0 & 0 & \xbar{1.000} & \xbar{1.000}\\
Push Down Field & 5 & 0.3 & 5 & 0 & 0 & \xbar{1.000} & \xbar{1.000}\\
Move Field & 1 & 0.5 & 1 & 1 & 0 & \xbar{0.500} & \xbar{1.000}\\
\addlinespace
Total & 102 & & 96 & 16 & 6 & \xbar{0.857} & \xbar{0.941}\\
\bottomrule
\end{tabular}
\end{table}

Table~\ref{TabCalibration} also presents the number of entries (\#) in the oracle for each refactoring relationship type, and the results we achieved using the optimal thresholds devised from the calibration process. The four rightmost columns show, respectively, the number of true positives (TP), the number of false positives (FP), the number of false negatives (FN), the precision and the recall.
In total, 85.7\% of the refactoring relationships reported by RefDiff were correct (precision) and it was able to find 94.1\% of the known refactorings (recall).

\section{Evaluation}
\label{SecEval}

\subsection{Precision and Recall}
\label{SecPrecRec}

\begin{table*}[t]
\renewcommand{\arraystretch}{1.2}
\setlength{\tabcolsep}{4pt}
\caption{Selected projects}
\label{TabProjects}
\centering
\begin{tabular}{@{}lp{10cm}r@{}}
\toprule
Repository URL & Description & LOC \\
\midrule
\url{github.com/Atmosphere/atmosphere}   &   The Asynchronous WebSocket/Comet Framework &   65,841\\
\url{github.com/clojure/clojure}   &  The Clojure programming language &   58,417\\
\url{github.com/google/guava}  &  Google Core Libraries for Java 6+ &   374,068\\
\url{github.com/dropwizard/metrics} &   Capturing JVM- and application-level metrics, so you know what's going on &   24,242 \\
\url{github.com/orientechnologies/orientdb} &   An Open Source NoSQL DBMS with the features of both Document and Graph DBMSs  &   168,924\\
\url{github.com/square/retrofit}  &   Type-safe HTTP client for Android and Java by Square, Inc. &   17,073\\
\url{github.com/spring-projects/spring-boot} &   Spring Boot makes it easy to create Spring-powered, production-grade applications and services with absolute minimum fuss  &  39,190 \\
\bottomrule
\end{tabular}

\end{table*}

To evaluate precision and recall, we compared the output of RefDiff with an oracle of known refactoring instances, similarly to the calibration procedure described in Section~\ref{SecCalibration}.
Besides, we compared our tool with three existing approaches, namely Refactoring Miner~\cite{fse2016-why-we-refactor}, Refactoring Crawler~\cite{dig2006automated}, and Ref-Finder~\cite{Kim:2010:RefFinder}.

\subsubsection{Construction of the oracle}
\label{SecOracleConstruction}

In the calibration procedure (Section~\ref{TabCalibration}), we used an oracle containing refactoring instances found on commits from open-source software repositories.
This strategy has the advantage of using real refactoring instances, but it also has a drawback.
There are no practical means of assuring that the oracle contains all existing refactoring instances in a given commit. In many cases, a single commit changes several files in non trivial ways, and a manual inspection of all changes using diff tools is time-consuming and error-prone.
Thus, refactoring instances might be missed by the tool under evaluation and also by the oracle.
Therefore, the computation of recall may not be reliable.

To be able to reliably compute recall, we employed the strategy of building an evaluation oracle by deliberately applying refactoring in software repositories in a controlled manner, similarly to Chaparro~et~al.~\cite{Chaparro:2014}. Such refactorings were applied by graduate students of a Software Architecture course. First, we randomly selected a list of 20 GitHub hosted Java repositories from the dataset of Silva et al.~\cite{fse2016-why-we-refactor} that contained a Maven project file (\texttt{pom.xml}). This way, we could use the Maven tool to build the project and import its source code to Eclipse IDE. Then, the professor of the course (an author of this paper) asked the students to:
\begin{enumerate}
\item Choose one of the 20 Java repositories in the list, given the constraint that a repository could not be taken by two students.
\item Analyze the latest revision of the source code, apply a specified number of refactoring operations on it, and commit the changes.
The students were instructed to apply at least three refactorings of each refactoring type listed in Table~\ref{TabOracle}.
\item Document all refactoring operations applied in a spreadsheet, using a specified format.
\end{enumerate}

\begin{table}[htbp]
\renewcommand{\arraystretch}{1.2}
\setlength{\tabcolsep}{4pt}
\caption{Refactoring types in the evaluation oracle}
\label{TabOracle}
\centering
\begin{tabular}{@{}lrrrrr@{}}
\toprule
          & & \multicolumn{4}{c}{Supported by}\\
\cmidrule(lr){3-6}
Ref. Type & \# & RDiff & RMinr & RCraw & RFind\\
\midrule
Rename Type & 35 & yes & yes & yes & no\\
Move Type & 31 & yes & yes & no & no\\
Extract Superclass & 16 & yes & yes & no & yes\\
Rename Method & 70 & yes & yes & yes & yes\\
Pull Up Method & 15 & yes & yes & yes & yes\\
Push Down Method & 68 & yes & yes & yes & yes\\
Move Method  & 31 & yes & yes & yes & yes\\
Extract Method & 29 & yes & yes & no & yes\\
Inline Method & 52 & yes & yes & no & yes\\
Pull Up Field & 33 & yes & yes & no & yes\\
Push Down Field & 42 & yes & yes & no & yes\\
Move Field  & 26 & yes & yes & no & yes\\
\addlinespace
Total  & 448 &  &  &  & \\
\bottomrule
\end{tabular}

\end{table}

It is worth noting that refactoring operations documented by them were confirmed by the first author of the paper by inspecting the source code. In this step, minor mistakes and typos were fixed.
For example, in some cases students typed the name of a class or method incorrectly. There were also a few cases of refactorings actually applied that were not reported in the spreadsheet. For example, a student inlined a method into the body of two other methods, but incorrectly reported only one of them in the spreadsheet.

By the end of the deadline, seven students properly completed the tasks and applied the refactorings in the repositories listed in Table~\ref{TabProjects}. Note that the repositories contain relevant Java projects such as Google Guava, Spring Boot and OrientDB. Table~\ref{TabProjects} also presents a short description of the project and the number of lines of Java code within each repository.

In total, we included 448 refactoring relationships in the evaluation oracle, as presented in Table~\ref{TabOracle}, covering 12 well-known refactoring types.
Note that a refactoring operation may be represented by more than one refactoring relationship. For example, a method $m$ may be extract from both $x$ and $y$. In this case, the oracle would contain the relationships $\mathit{ExtractMethod}(x, m)$ and $\mathit{ExtractMethod}(y, m)$.


\subsubsection{Execution of the selected approaches}

\begin{table*}[t]
\renewcommand{\arraystretch}{1.2}
\setlength{\tabcolsep}{4pt}
\caption{Precision and recall by refactoring type}
\label{TabEvaluationRt}
\centering
\begin{tabular}{@{}lcccccccc@{}}
\toprule
          & \multicolumn{2}{c}{RDiff} & \multicolumn{2}{c}{RMinr} & \multicolumn{2}{c}{RCraw} & \multicolumn{2}{c}{RFind}\\
\cmidrule(lr){2-3} \cmidrule(lr){4-5} \cmidrule(lr){6-7} \cmidrule(lr){8-9}
Ref. Type & Precision & Recall & Precision & Recall & Precision & Recall & Precision & Recall\\
\midrule
Rename Type & \xbar{1.000} & \xbar{1.000} & \xbar{1.000} & \xbar{1.000} & \xbar{0.750} & \xbar{0.429} &              &             \\
Move Type & \xbar{1.000} & \xbar{0.968} & \xbar{1.000} & \xbar{0.968} &              &              &              &             \\
Extract Superclass & \xbar{1.000} & \xbar{0.875} & \xbar{1.000} & \xbar{0.875} &              &              & \xbar{0.484} & \xbar{0.938}\\
Rename Method & \xbar{1.000} & \xbar{0.943} & \xbar{1.000} & \xbar{0.886} & \xbar{0.971} & \xbar{0.486} & \xbar{0.868} & \xbar{0.843}\\
Pull Up Method & \xbar{1.000} & \xbar{0.600} & \xbar{1.000} & \xbar{0.733} & \xbar{0.500} & \xbar{0.067} & \xbar{1.000} & \xbar{0.571}\\
Push Down Method & \xbar{1.000} & \xbar{0.971} & \xbar{1.000} & \xbar{0.176} & \xbar{1.000} & \xbar{0.265} & \xbar{1.000} & \xbar{0.491}\\
Move Method  & \xbar{1.000} & \xbar{1.000} & \xbar{1.000} & \xbar{0.742} & \xbar{0.090} & \xbar{0.323} & \xbar{0.054} & \xbar{0.759}\\
Extract Method & \xbar{1.000} & \xbar{0.897} & \xbar{1.000} & \xbar{0.862} &              &              & \xbar{0.607} & \xbar{0.586}\\
Inline Method & \xbar{1.000} & \xbar{0.981} & \xbar{1.000} & \xbar{0.423} &              &              & \xbar{0.917} & \xbar{0.688}\\
Pull Up Field & \xbar{1.000} & \xbar{0.576} & \xbar{1.000} & \xbar{0.970} &              &              & \xbar{1.000} & \xbar{0.394}\\
Push Down Field & \xbar{1.000} & \xbar{0.929} & \xbar{1.000} & \xbar{0.929} &              &              & \xbar{1.000} & \xbar{0.333}\\
Move Field  & \xbar{1.000} & \xbar{0.269} & \xbar{0.583} & \xbar{0.808} &              &              & \xbar{0.097} & \xbar{0.923}\\
\bottomrule
\end{tabular}

\end{table*}

After constructing the evaluation oracle, we run RefDiff, Refactoring Miner (1.0.0), Refactoring Crawler (1.0.0), and Ref-Finder (1.0.4) to compare their output with the refactoring relationships in the oracle.
Refactoring Miner can be used as an API, and it provides mechanisms to export its output, thus, we only needed to transform it into a normalized format.
Refactoring Miner and Refactoring Crawler are plug-ins that depend on Eclipse IDE. In both cases, we needed to adapt their source code to enable or to facilitate the evaluation.
We faced an issue to run Refactoring Crawler on the selected projects, because it was dependent on an outdated version of the Eclipse IDE, in which we were unable to import the projects through the Maven-Eclipse integration. To resolve such issue, we decided to adapt the source code of Refactoring Crawler to a recent Eclipse release (Mars). The necessary code modifications were simple, but, as a precaution, we assessed whether the results of our modified version of the tool were identical to those achieved by the original implementation using the evaluation dataset provided by the authors.
In the case of Ref-Finder, we also had to modify its source code, but the reason was to be able to export its output into a text file.

Another issue we faced with Ref-Finder was related with refactorings that involved methods, because Ref-Finder only displays the name of the method and its class, but not its complete signature. Therefore, when there are overloaded methods in a class, Ref-Finder's output is ambiguous. To overcome this issue, we adopted a less strict check that ignores method parameters when comparing Ref-Finder's output with entries in the oracle. For example, if that the oracle contains the entry:\\[-10pt]

\noindent$\mathit{ExtractMethod}(\mathtt{Calc.mult(int, int)}, \mathtt{Calc.add(int, int)})$\\[-10pt]

\noindent and Ref-Finder reports:\\[-10pt]

\noindent$\mathit{ExtractMethod}(\mathtt{Calc.mult}, \mathtt{Calc.add})$\\[-10pt]

\noindent we still consider it a true positive.


Last, it is worth noting that the refactoring types contained in the oracle are not supported by all approaches, as detailed in Table~\ref{TabOracle}. 
For example, Refactoring Crawler does not support the detection of \emph{Move Attribute} refactorings. We decided to disregarded such entries of the oracle when counting the number of false negatives. This means that an approach may achieve 1.0 recall even if it does not support all refactoring types in the oracle.

\subsubsection{Results and discussion}

The overall precision and recall for each approach are presented in Table~\ref{TabEvaluationOverall}.
RefDiff achieves the best precision (1.000), followed by Refactoring Miner (0.956), Refactoring Crawler (0.419), and Ref-Finder (0.264).
In terms of recall, RefDiff still holds the best result (0.877), followed by Refactoring Miner (0.728), Ref-Finder (0.642), and Refactoring Crawler (0.356).

\begin{table}[h]
\renewcommand{\arraystretch}{1.2}
\setlength{\tabcolsep}{4pt}
\caption{Overall precision and recall}
\label{TabEvaluationOverall}
\centering
\begin{tabular}{@{}lrrrll@{}}
\toprule
Approach & TP & FP & FN & Precision & Recall\\
\midrule
RDiff & 393 & 0 & 55 & \xbar{1.000} & \xbar{0.877} \\
RMinr & 326 & 15 & 122 & \xbar{0.956} & \xbar{0.728} \\
RCraw & 78 & 108 & 141 & \xbar{0.419} & \xbar{0.356} \\
RFind & 231 & 645 & 129 & \xbar{0.264} & \xbar{0.642} \\
\addlinespace
RCraw* & 78 & 56 & 141 & \xbar{0.582} & \xbar{0.356} \\
RFind* & 231 & 241 & 129 & \xbar{0.489} & \xbar{0.642} \\
\bottomrule
\end{tabular}

\end{table}

Detailed precision and recall results for each refactoring type are presented in Table~\ref{TabEvaluationRt}. 
We can note that the results for some refactoring types stand out from the rest. For example, RefDiff achieved a recall of only 0.269 for \emph{Move Field}. This observation suggests that the threshold for such refactoring could possibly be less restrictive. For Refactoring Miner, the main offender in terms of recall is \emph{Push Down Method}.

\begin{table*}[t]
\renewcommand{\arraystretch}{1.2}
\setlength{\tabcolsep}{4pt}
\caption{Execution time}
\label{TabExecTime}
\centering
\begin{tabular}{@{}lrrrrrrrrrr@{}}
\toprule
 &  & \multicolumn{4}{c}{RDiff execution time} & \multicolumn{4}{c}{RMinr execution time}\\
\cmidrule(lr){3-6} \cmidrule(lr){7-10}
Repository & Commits & Min. (ms) & Max. (ms) & Avg. (ms) & Total. (s) & Min. (ms) & Max. (ms) & Avg. (ms) & Total. (s)\\
\midrule
\url{androidannotations/androidannotations} & 29 & 1 & 4,956 & 451 & 13 & 1 & 1,753 & 211 & 6\\
\url{bumptech/glide} & 41 & 1 & 3,349 & 594 & 24 & 2 & 8,992 & 466 & 19\\
\url{elastic/elasticsearch} & 946 & 1 & 42,344 & 1,897 & 1,795 & 1 & 103,943 & 1,105 & 1,046\\
\url{libgdx/libgdx} & 69 & 0 & 5,112 & 805 & 56 & 1 & 6,774 & 578 & 40\\
\url{netty/netty} & 225 & 0 & 3,384 & 640 & 144 & 0 & 59,736 & 665 & 150\\
\url{PhilJay/MPAndroidChart} & 14 & 1 & 816 & 245 & 3 & 1 & 310 & 79 & 1\\
\url{ReactiveX/RxJava} & 120 & 1 & 810,744 & 10,475 & 1,257 & 1 & 17,369 & 538 & 65\\
\url{spring-projects/spring-framework} & 478 & 1 & 15,019 & 1,205 & 576 & 1 & 6,133 & 920 & 440\\
\url{square/okhttp} & 45 & 1 & 1,526 & 380 & 17 & 1 & 616 & 178 & 8\\
\url{zxing/zxing} & 23 & 1 & 773 & 342 & 8 & 1 & 502 & 230 & 5\\
\addlinespace
Total & 1990 & 0 & 810,744 & 1,956 & 3,893 & 0 & 103,943 & 894 & 1,779\\
\bottomrule
\end{tabular}

\end{table*}

When we focus on Refactoring Crawler and Ref-Finder, one fact that clearly draws one's attention is the extremely low precision for \emph{Move Method} and \emph{Move Field}. A more detailed analysis revealed that one reason for this was the lack of \emph{Move Type} and/or \emph{Rename Type} detection support in these approaches. For example, in the case of a class $A$ is moved to become $A'$, several \emph{Move Method} and \emph{Move Field} relationships from members of class $A$ to class $A'$ are mistakenly reported.
This issue drastically affects the precision of such approaches. For example, 284 (74\%) out of 382 \emph{Move Method} false positives reported by Ref-Finder are due to this reason. Thus, we decided to recompute the overall precision for Refactoring Crawler and Ref-Finder disregarding false positives that fell in that scenario. The last two lines of Table~\ref{TabEvaluationOverall} presents the recomputed results for both tools, under the names of \emph{RCraw*} and \emph{RFind*}. We can note a significant improvement in precision for Refactoring Crawler (from 0.419 to 0.582) and Ref-Finder (from 0.264 to 0.489).
However, even in this scenario, RefDiff and Refactoring Miner results are still far ahead from them. We should also note that our results corroborate with the findings of Soares~et~al.~\cite{Soares:2013} in a study with JHotDraw and Apache Common Collections, in which Ref-Finder achieved precision of 0.35 and recall of 0.24.

It is interesting to note that the precision achieved in the calibration process was inferior to the one achieved in the evaluation, specially considering that the thresholds were optimized to that data.
However, such behavior is not surprising, because the calibration oracle is composed of real refactorings performed in those systems, possibly interleaved with all kinds of code changes. Such scenario is undoubtedly more challenging for refactoring detection tools than refactoring-only commits.

\subsection{Execution Time}

Besides evaluating precision and recall, we also designed a study to evaluate the execution time and scalability of RefDiff and Refactoring Miner, in the context of mining refactorings from open-source software repositories.
Ref-Finder and Refactoring Crawler were removed from the comparison because they are Eclipse-based plug-ins, which are not suitable for automation. Specifically, there are two issues that hinder their application. First, they require user interaction through the Eclipse UI to select their input and trigger the refactoring detection. Second, they require each pair of versions under comparison to be imported and configured as Eclipse projects. Thus, such tasks cannot be reliably automated. In contrast, both RefDiff and Refactoring Miner are able to detect refactorings directly from commits in git repositories, comparing the revisions of the source code before and after the changes.

To run the study, we selected the ten most popular Java repositories from GitHub that met the following criteria:
(i) the repository was not used in the studies from Section~\ref{SecPrecRec} and Section~\ref{TabCalibration};
(ii) the repository contains a real software component (i.e., not a toy example or tutorial);
(iii) the repository contains at least 1,000 commits; and
(iv) the repository contains commits pushed in the last three months.
Then, we employed RefDiff and Refactoring Miner to analyze each commit found in the default branch of the repositories, in a time window ranging from January 1, 2017 to March 27, 2017. For simplicity, merge commits were excluded from the analysis, as the code changes they contain are usually devised from other commits.

\begin{figure}[b]
\centering
\includegraphics[width=0.45\linewidth]{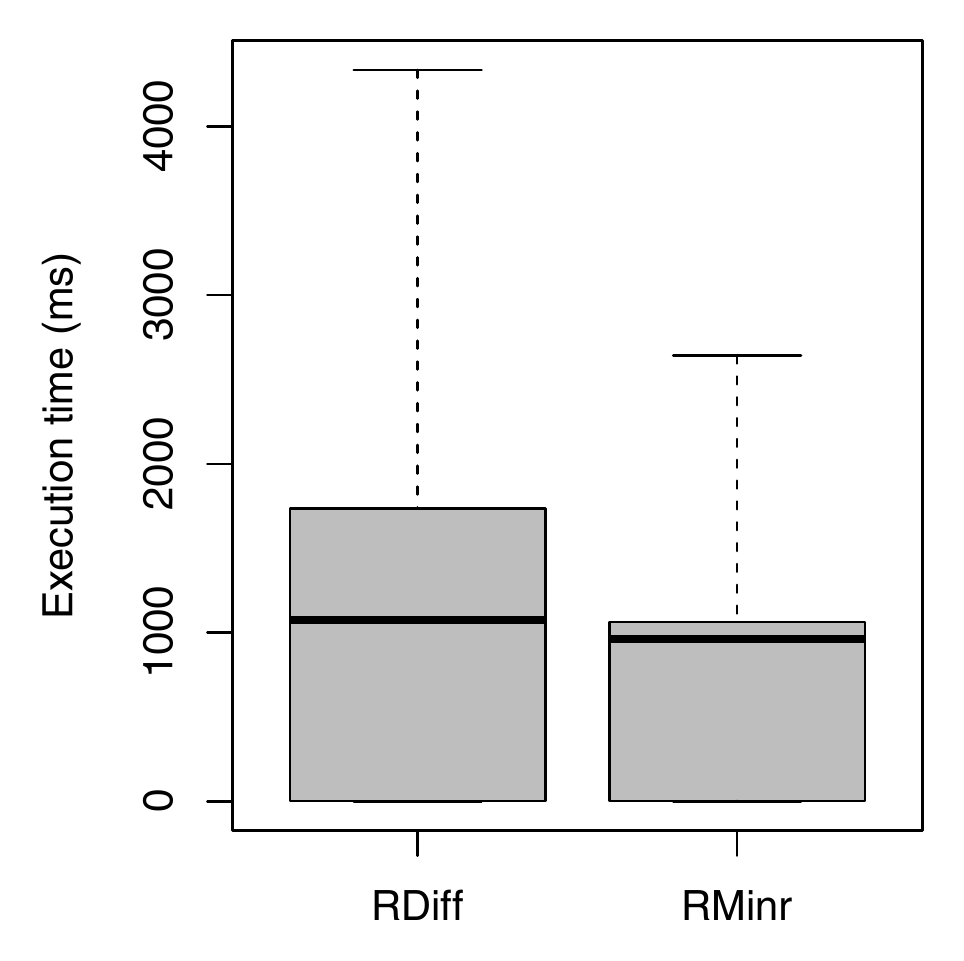}
\caption{Boxplot of execution time per commit (ommiting outliers)}
\label{FigBoxplot}
\end{figure}

Table~\ref{TabExecTime} shows the selected repositories, the number of analyzed commits, the execution time (minimum, maximum, and average) per commit, and the total execution time, for each approach.
On average, RefDiff spends 1.96 seconds to detect refactorings in a commit, while Refactoring Miner spends 0.89 seconds. The total execution time to analyze the data set was 3,893 seconds (about 64 min) for RefDiff, against 1,779 seconds (about 29 min) for Refactoring Miner. In the worst case, RefDiff spent 810 seconds (about 13 minutes) in single commit. However, such case happened only in a commit from \url{ReactiveX/RxJava}. For all other repositories, the worst execution time was less than a minute.
Figure~\ref{FigBoxplot} shows a box plot of the distribution of execution time per commit (omitting outliers for readability). We can note that the median is similar for both approaches (close to one second), with a slight advantage to Refactoring Miner.
It is clear from the results that Refactoring Miner achieves lower execution time in most cases, but such differences are relatively small in practice. Thus, the potential gain in precision and recall using RefDiff may still be worth it. Besides, the implementation of RefDiff had as its main objective the evaluation of the approach. Thus, it is possible that its source code can be optimized.
In summary, we can conclude that both approaches provide acceptable performance and scalability, enabling their application in large code bases, such as Elasticsearch (922~KLOC) and Spring Framework (1,016~KLOC).

\section{Threats to Validity}
\label{SecThreats}

\noindent\textbf{External validity}: 
The evaluation of precision and recall of RefDiff used refactoring instances injected in  seven popular open-source Java projects. We cannot claim that the precision and recall of our approach would be the same for different projects, with distinct characteristics, and with actual refactorings applied by developers. However, such setup was necessary to compute recall, as discussed in Section~\ref{SecOracleConstruction}. Besides, the results we achieved in the calibration process (precision of 85.7\% and recall of 94.1\%), in which we used actual commits from relevant Java repositories, suggest that RefDiff's precision and recall are acceptable in real scenarios. Nevertheless, we plan to extend this study by assessing the precision of RefDiff in a large corpus of commits from open-source repositories.


\noindent\textbf{Internal validity}: The evaluation oracle we used in our study is subject to human errors due to the manual task of applying the refactorings and documenting them. However, we addressed that issue by inspecting the source code of the refactored systems to validate all documented refactorings before running our experiment.
Besides, the procedures to compare the output of each tool with the entries in the oracle were automated to avoid any mistake.

\section{Conclusion}
\label{SecConclusion}

In this paper, we propose RefDiff, an approach to detect refactorings in version histories of software components.
Our approach employs a combination of heuristics based on static analysis and code similarity to detect 13 well-known refactoring types.
One key aspect of our algorithm is the employed similarity index, which is an adaptation of the TF-IDF weighting scheme.
We have also evaluated RefDiff, comparing it with Refactoring Miner, Refactoring Crawler, and Ref-Finder, using on oracle of 448 known refactorings across seven Java projects. RefDiff achieved the best result among the evaluated tools: precision of 1.00 and recall of 0.88.
As an additional contribution, we made publicly available the implementation of RefDiff and all data used in the experiments.

As future work, we intend to explore applications of {RefDiff}. For example, we could use information of refactorings to build an improved code diff visualization that presents changes in refactored code elements side-by-side with their matching elements in the previous version.
Besides, a reliable refactoring detection tool open up possibilities for novel empirical studies on refactoring practices, taking advantage of the vast amount of historical information available in code repositories.

\section*{Acknowledgment}

Our research is supported by FAPEMIG and CNPq.



\bibliographystyle{IEEEtran}
\bibliography{IEEEabrv,references}
%



\end{document}